\documentclass[11pt, a4paper]{article}

\usepackage[margin=1in]{geometry}
\usepackage{graphicx}
\graphicspath{ {./figures/} }
\usepackage{setspace}
\usepackage{hyperref}
\usepackage{caption}
\usepackage{amsfonts}
\usepackage{amsmath}
\usepackage{bm}
\usepackage{booktabs}
\usepackage{float}
\usepackage{authblk}

\usepackage[
backend=biber,
style=alphabetic,
sorting=ynt
]{biblatex}

\title{\textbf{Online FDR Control for RNAseq Data}}

\author[a]{Lathan Liou}
\author[a]{Milena Hornburg}
\author[b]{David S.\ Robertson}

\affil[a]{Merck \& Co., Inc., Kenilworth, NJ, USA}

\affil[b]{MRC Biostatistics Unit, School of Clinical Medicine, University of Cambridge, UK}

\begin{document}

\maketitle

\onehalfspace

\begin{abstract}
Motivation: While the analysis of a single RNA sequencing (RNAseq) dataset has been well described in the literature, modern research workflows often have additional complexity in that related RNAseq experiments are performed sequentially over time. The simplest and most widely used analysis strategy ignores the temporal aspects and analyses each dataset separately. However, this can lead to substantial inflation of the overall false discovery rate (FDR). We propose applying recently developed methodology for online hypothesis testing to analyse sequential RNAseq experiments in a principled way, guaranteeing FDR control at all times while never changing past decisions. 

Results: We show that standard offline approaches have variable control of FDR of related RNAseq experiments over time and a naively composed approach may improperly change historic decisions. We demonstrate that the online FDR algorithms are a principled way to guarantee control of FDR. Furthermore, in certain simulation scenarios, we observe empirically that online approaches have comparable power to offline approaches. 

Availability and Implementation: The onlineFDR package is freely available at \url{http://www.bioconductor.org/packages/onlineFDR}. Additional code used for the simulation studies can be found at \url{https://github.com/latlio/onlinefdr_rnaseq_simulation}

Contact: \href{mailto:david.robertson@mrc-bsu.cam.ac.uk}{david.robertson@mrc-bsu.cam.ac.uk} 
\\

\end{abstract}


\section{Introduction}

RNA sequencing (RNAseq) is a powerful and widely used tool to profile the expression of thousands of transcripts or genes in parallel, allowing scientists to study a biological system at the transcriptome level. Most RNAseq experiments aim to compare two biological states such as a treated and untreated system to better understand the biological differences between the two states. At the core of this analysis is the differential expression analysis that compares the expression of genes between those two states. Since hundreds to thousands of genes are simultaneously tested in high-throughput experiments for their differential expression between two biological states, multiple hypothesis testing correction is essential to control the rate of type~I errors (i.e.\ a false positive differential expression result). Earlier methods aimed to control a metric known as the family-wise error rate (FWER), but it is often highly conservative, controlling the probability of any false positives at the cost of greatly reduced power to detect true positives \cite{korthauer_practical_2019}. The false discovery rate (FDR), or expected proportion of discoveries which are falsely rejected \cite{benjamini_controlling_1995}, was more recently proposed as an alternative metric to the FWER in multiple testing control. FDR control has become a largely accepted standard in genomics \cite{chen_false_2010}. Standard RNAseq analysis approaches like DESeq2 and limma typically rely on the classic Benjamini–Hochberg procedure to control the FDR \cite{love_moderated_2014} \cite{ritchie_limma_2015}. 

While the analysis of a \textit{single} RNAseq dataset has been well described in the literature, the widespread application of sequencing in modern research often results in a temporal accumulation of potentially related RNAseq datasets. If datasets have the same comparison groups of interest and sufficient metadata to adjust for any confounding differences in experimental or study design, they may be combined in some way to pool information. The simplest and most widely used analysis strategy is to just ignore the temporal aspect and analyse each RNAseq dataset separately. However, this can lead to substantial inflation of the overall FDR~\cite{zrnic_power_2021}. Intuitively, we want to control FDR over all the RNAseq datasets that are related in some way (e.g. transcriptome of the same biological population). Open-source repositories of high-throughput gene expression data such as Gene Expression Omnibus are examples of growing datasets where similar RNAseq datasets are collected \cite{edgar_gene_2002}.

An alternative analysis strategy is to pool together the differential expression results of the RNAseq datasets as they become available, and then apply classical methods of FDR control on the \emph{p}-values. However, the critical disadvantage of this approach is that the decisions (i.e. the hypotheses rejected) for a particular dataset may change over time as new datasets are added, which is clearly undesirable in most application areas (see Section 3).

In light of these limitations, we propose applying recently developed methodology for \textit{online} hypothesis testing to analyse sequential RNAseq experiments in a principled way, guaranteeing FDR control at all times while never changing past decisions. In the framework of online hypothesis testing, the investigator must decide whether to reject the current null hypothesis without knowing the future $p$-values (or even the total number of hypotheses to be tested), but only knowing the historic decisions to date. Recent proposals have extended this framework to allow \textit{batches} of hypotheses to be tested in this way, which we exploit in our context of analysing RNAseq datasets. Batches are simply sets of \emph{p}-values; for instance, a differential expression analysis of one RNAseq dataset generates a batch of \emph{p}-values, and the analysis of another RNAseq dataset generates another batch of \emph{p}-values. Robertson et al. describe scenarios in which online hypothesis testing can be applied to growing public biological databases such as the International Mouse Phenotype Consortium \cite{robertson_onlinefdr_2019}. 

Through a simulation study of RNAseq datasets, we compare and contrast the aforementioned three analysis strategies, showing that the online hypothesis testing algorithms indeed control the FDR. Furthermore, in certain simulation scenarios, we observe empirically that online approaches can even be competitive with the other approaches in terms of power. We also illustrate the proposed methodology on three real-world RNAseq datasets arriving over time.

\section{FDR Control Paradigms}

In this paper, we compare three different FDR control methods and distinguish between “offline” and “online” methods (Figure 1). We also want to delineate the vocabulary we use in this paper: ``dataset" or ``batch" refers to a set of differential expression \emph{p}-values (note that this is different from control or experimental ``samples" in a standard two-group differential expression comparison experiment); ``family" refers to a group of datasets. 

\begin{figure}[H]
\centering
\includegraphics[scale=0.5]{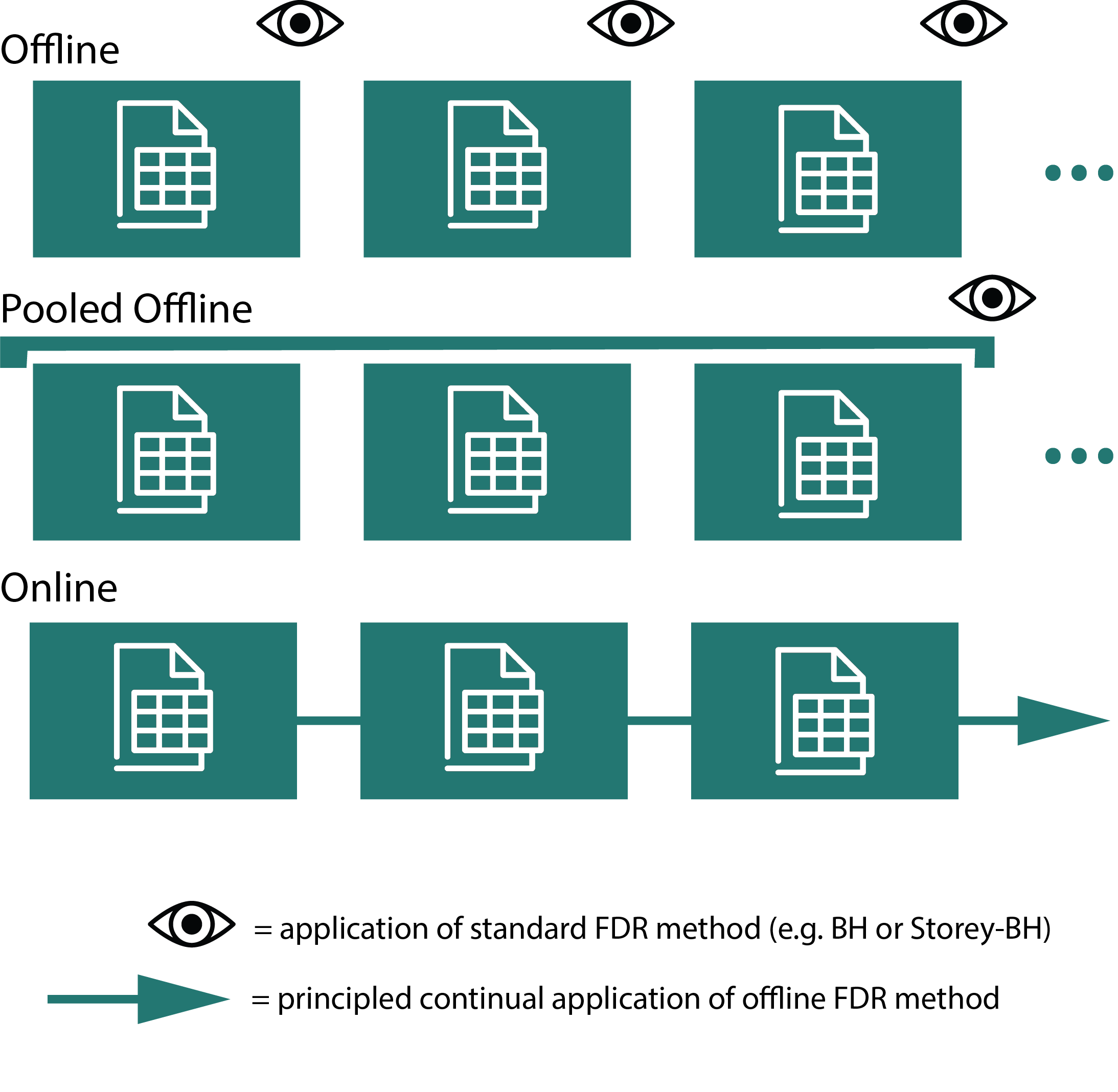}
\caption{Diagram of offline, pooled offline, and online FDR control approaches}
\end{figure}

We use the term ``offline" to refer to methods that take as input a single set of data and output a set of rejection decisions for all hypotheses at once to control the FDR. The offline approach assumes no knowledge of previous or future datasets and only considers the dataset obtained at any given time (Figure 1). This approach is what is widely performed in academia and industry; we note that ``offline" is not a common designation, but we describe this conventional approach as such to distinguish it from ``online" method, which we describe later.

In our paper, for the offline approaches, we use the Benjamini-Hochberg (BH) procedure \cite{benjamini_controlling_1995} and Storey-BH (StBH) procedure \cite{storey_direct_2002}. Given p-values $p_1, p_2, \ldots, p_N$ and a significance level $\alpha$, the BH procedure is as follows:

\begin{enumerate}
    \item Order the \emph{p-}values from smallest to largest, giving ordered \emph{p-}values $p_{(1)} \leq \cdots \leq p_{(N)}$ and define $p_{(0)} = 0$;
    \item Let $i^*$ be the maximal index such that $p_{(i^*)} \leq i\alpha/N$;
    \item Reject null hypothesis $H_j$ for every test with $p_j \leq p_{(i^*)}$.
\end{enumerate}

StBH improves upon the original BH procedure by letting users choose a parameter $\lambda \in (0,1)$, which is used to estimate the proportion of nulls in the \emph{p-}value set, defined as: 

$$\frac{1+\Sigma^N_{i=1} \bm{1} \{ p_i > \lambda \}}{N(1-\lambda)}$$

The motivation for using StBH is that BH might be overly conservative when there are many expected rejections with a strong signal, so StBH is more adaptive to the set of \emph{p-}values at hand, which will provide important intuition to the online analogues of these algorithms. It is important to note that BH relies on the assumption that the \emph{p-values} are positively dependent \cite{yoav_benjamini_control_2001} and StBH requires independent \emph{p}-values \cite{storey_direct_2002}.

We use the term ``pooled offline" to describe an extension of the classical offline approach in which multiple datasets are appended into a single pooled dataset and classical FDR control methods (e.g. BH are then applied to the pooled dataset; Figure 1). The pooled offline approach may be expected, at first glance, to properly control FDR across all datasets since it takes into consideration previous hypotheses from prior datasets; however, we demonstrate later that this is not the case. 

To illustrate the pooled offline approach, let's consider a toy example where we have 5 datasets, which have arrived over time, each containing $N$ hypotheses. Each dataset $d$ contains $N$ \emph{p-}values $p_{d1}, p_{d2}, \ldots, p_{dN}$. If we applied a pooled offline approach at the time the fifth dataset arrived, we combine the 5 sets of \emph{p-}values $p_{11}, p_{12}, \ldots, p_{21}, \ldots, p_{51}, \ldots p_{5N}$ and then apply either the BH or Storey-BH procedure as described above. Note that the pooled offline approach applied to a single dataset is exactly equivalent to the offline approach. The important assumption with offline methods is that they require knowledge of all the \emph{p-}values, which does not address the situation where we do not know the total number of hypotheses or future \emph{p-}values.

In contrast, the online approach allows for informed hypothesis-driven decision-making using prior datasets and offers guaranteed FDR control for potential future datasets. With offline methods, there are distinct points in time when an FDR control method is applied, but with online methods, the application is continuous throughout the timeline of datasets: past, present, and future (Figure 1). We describe the online paradigm for testing batches of $N$ hypotheses arriving sequentially over time, where each batch represents a single RNAseq experiment. At each time $t \in \mathbb{N}$, a batch of $N_t$ hypotheses arrives. In the online paradigm, no information about future batches needs to be available, such as how many hypotheses each batch contains or the total number of batches. We denote the set of hypotheses in the \emph{t-}th batch by $\bm{H_t} := \{ H_{t,1}, \ldots ,H_{t,N_t} \} $. Each hypothesis has a $p$-value associated with it. Let $\bm{P_t}$ denote the set of \emph{p-}values corresponding to the \emph{t-}th batch of hypotheses, given by 
$\bm{P_t} := \{ P_{t,1}, \ldots ,P_{t,N_t} \}$, where $P_{t,j}$ is the \emph{j-}th \emph{p-}value in batch $t$. The FDR \emph{up to the time t} is then the expected false discovery proportion (FDP) up to time $t$, i.e.\ the expected proportion of false discoveries up to time $t$ among the rejected hypotheses up to time $t$, for all batches $s$ up until time $t$. Letting $V(t)$ denote the number of incorrectly rejected hypotheses in batch~$t$ and $R(t)$ denote the number of rejected hypotheses in batch~$t$, the FDR is defined as follows:  

\begin{gather*}
\text{FDR}(t) \equiv \mathbb{E}[\text{FDP}(t)] \equiv \mathbb{E} \left( \frac{\sum^t_{s=1} V(s)}{\max(\sum^t_{s=1} R(s), 1)} \right) .
\end{gather*}

The batch procedures for online control FDR were originally described by Zrnic et al.\ (2021)~\cite{zrnic_power_2021} and implemented in the R package \emph{onlineFDR} \cite{robertson_onlinefdr_2019} which is freely available through Bioconductor at \url{http://www.bioconductor.org/packages/onlineFDR}.

In brief, batch procedures aim to interpolate between the offline and online settings and achieve high power, while guaranteeing FDR control at all times $t \in \mathbb{N}$. Batch algorithms adaptively determine a \emph{test level} $\alpha_t$ based on information about past batches of hypothesis tests. We use three Batch procedures: BatchBH, BatchStBH, and BatchPRDS. BatchBH and BatchStBH are the online analogues of the offline BH procedure and Storey-BH procedure respectively. Both BatchBH and BatchStBH assume that \emph{p}-values are independent. BatchStBH aims to be more adaptive to the data than BatchBH, given its requirement for a user-chosen constant $\lambda \in (0,1)$. BatchPRDS is a special case of BatchBH that controls the FDR when the \emph{p-}values in a single batch are positively dependent yet independent across batches. Further details about these algorithms can be found in the original paper \cite{zrnic_power_2021}. \\

\section{Real World Data Application}

We used three real-world datasets procured from internal studies of Merck \& Co., Inc., Kenilworth, NJ, USA. Datasets are anonymized as Batch 1, 2, and 3. All three datasets were RNAseq studies of CT26 syngeneic mouse models treated with murinized rat anti-mouse PD-1 antibody (muDX400) or vehicle control with the same experimental design such as treatment duration and dose. The order of datasets was determined by the chronological order that they were performed over time at Merck \& Co., Inc., Kenilworth, NJ, USA. We selected vehicle control samples as our control arm and muDX400 samples as our treatment arm in a two-sample differential expression analysis. Batch 1 contained 10 control and 10 treatment samples. Batch 2 contained 10 control and 5 treatment samples. Batch 3 contained 19 control and 18 treatment samples. After filtering out ambiguously mapped genes and genes with low counts across samples using the \emph{edgeR} package \cite{chen_reads_2016}, raw count data was voom-transformed and differential expression analysis was performed using the \emph{limma-voom} workflow \cite{ritchie_limma_2015}\cite{law_voom_2014}. 14708, 12230, and 14948 genes went into the differential expression analysis for Batch 1, 2, and 3 respectively. 

We note that in our two-sample differential expression analysis design for this study, we do not aim to identify biologically meaningful insights, but rather use a  vehicle control versus muDX400 treatment as a way to showcase the behavior of online FDR control methods compared to their offline counterparts. 

We apply offline, pooled offline, and three Batch online approaches to three real-world datasets (annotated as Batches 1, 2 and 3). As a benchmark, we see that almost all of the FDR control methods are more conservative than the uncorrected approach at an $\alpha$ of 0.05 at all time points (Table 1). The offline StBH and pooled offline StBH approaches make more rejections compared to their respective BH counterparts. This is expected because StBH is more adaptive to the data at hand and has been shown empirically to yield higher power \cite{storey_direct_2002}. Zrnic et al. have shown theoretical guarantees that FDR will be controlled across all batches and even future batches, assuming \emph{p}-values are independent (\cite{zrnic_power_2021}. Although BatchBH and BatchStBH make fewer rejections than their offline and pooled offline analogues, it is likely that they are acting more conservatively to maintain overall FDR control. It is worth noting that in this real-world example, BatchStBH rejecte more genes as differentially expressed (DE) than offline and pooled BH at each time point (Table 1).

\begin{table}[H] 
\centering
\caption{Cumulative number of rejections (proportion of total genes tested) in real-world RNAseq data across batches. N is the total number of genes tested. Blue color scales with the number of null hypothesis rejections with dark blue for high number of rejections, light blue for low number of rejections.}
\includegraphics[width=\textwidth]{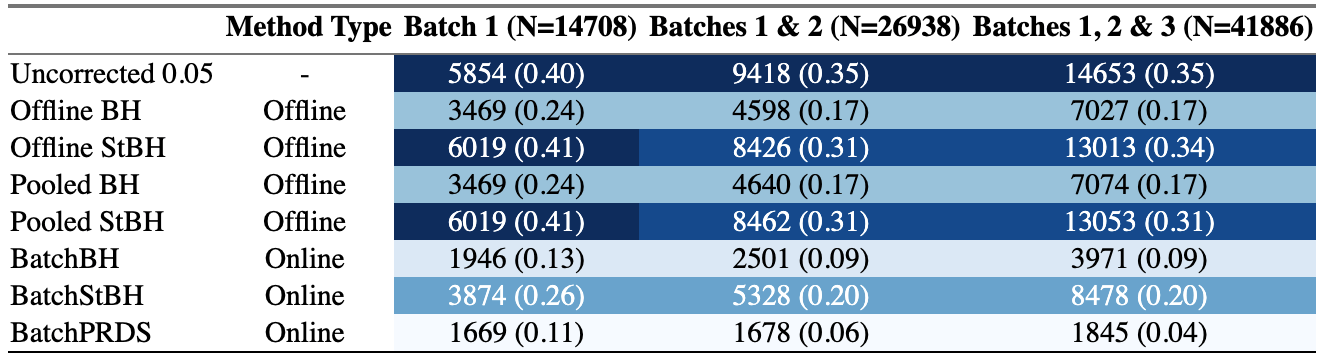}
\end{table}

The BatchBH procedure made the same rejections for DE genes 56\% of the time as its offline and pooled analogues across three accumulated datasets (Figure 2). BatchStBH achieved a higher proportion of overlapping rejections of 66\% across the three datasets. Online approaches are able to detect a moderately high percentage of the same DE genes as the offline approaches, but the online approaches in this real-world data example appear more conservative.

\begin{figure}[H]
\includegraphics[width=\textwidth]{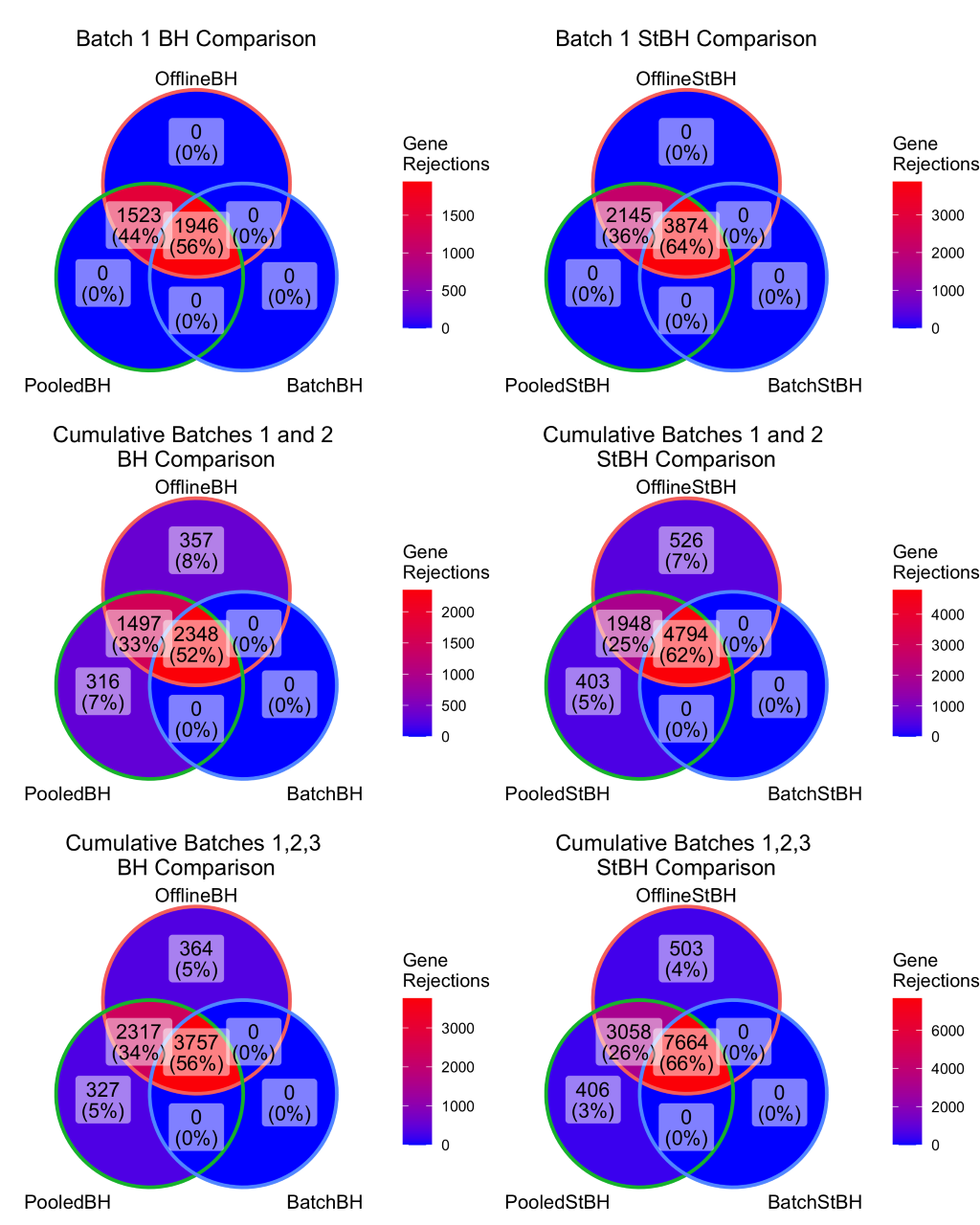}
\caption{Overlap in differentially expressed Genes Between Offline, Pooled Offline, and Batch Approaches, $\alpha = 0.05$}
\end{figure}

We next want to illustrate that naively composing offline FDR control approaches is not appropriate, using Pooled BH between Batches 1 and 2 as an example. In Batch 1, a set of 3469 genes were declared DE by Pooled BH (Table 1). In the pooled approach, we first performed the differential expression analysis independently for Batch 1 and Batch 2, and then appended the resulting p-values together and performed an FDR correction. Cumulatively between Batches 1 and 2, 4600 genes were declared DE; however, only 3069 genes from dataset 1 were declared DE. 400 genes ``lost" their significance because BH reorders all the \emph{p-}values. To further illustrate the change in historic decisions in a pooled BH approach, we randomly sampled 50 genes declared DE and 50 genes declared non-DE from Batch 1 100 times. Of the genes that were declared DE in Batch 1, the number of genes that had decisions changed with a pooled application ranged from 1 to 10 (average of 6). One of the random samples is shown in Figure 3. When comparing the differential expression decisions (e.g. null hypothesis rejections) for the sampled genes in Batch 1 to the decisions for the same sampled genes from Batch 1 but from the pooled FDR correction across Batches 1 and 2 and across Batches 1-3, we observe how the differential expression rejection decisions for the sampled genes from Batch 1 have changed. It is likely that \emph{p-}values that are close to the critical value as defined in the BH algorithm will have their decisions changed. Potentially changing whether a gene is rejected as being DE each time an additional batch of genes is tested is clearly infeasible from the practical perspective of wanting to act in a timely manner on the hypothesis test results (e.g.\ carrying out subsequent functional validation experiments). It is important to emphasize that online methods do not change the historic rejection decisions made when new datasets of \emph{p}-values arrive and are pooled with existing datasets. 

\begin{figure}[H]
\includegraphics[width=\textwidth]{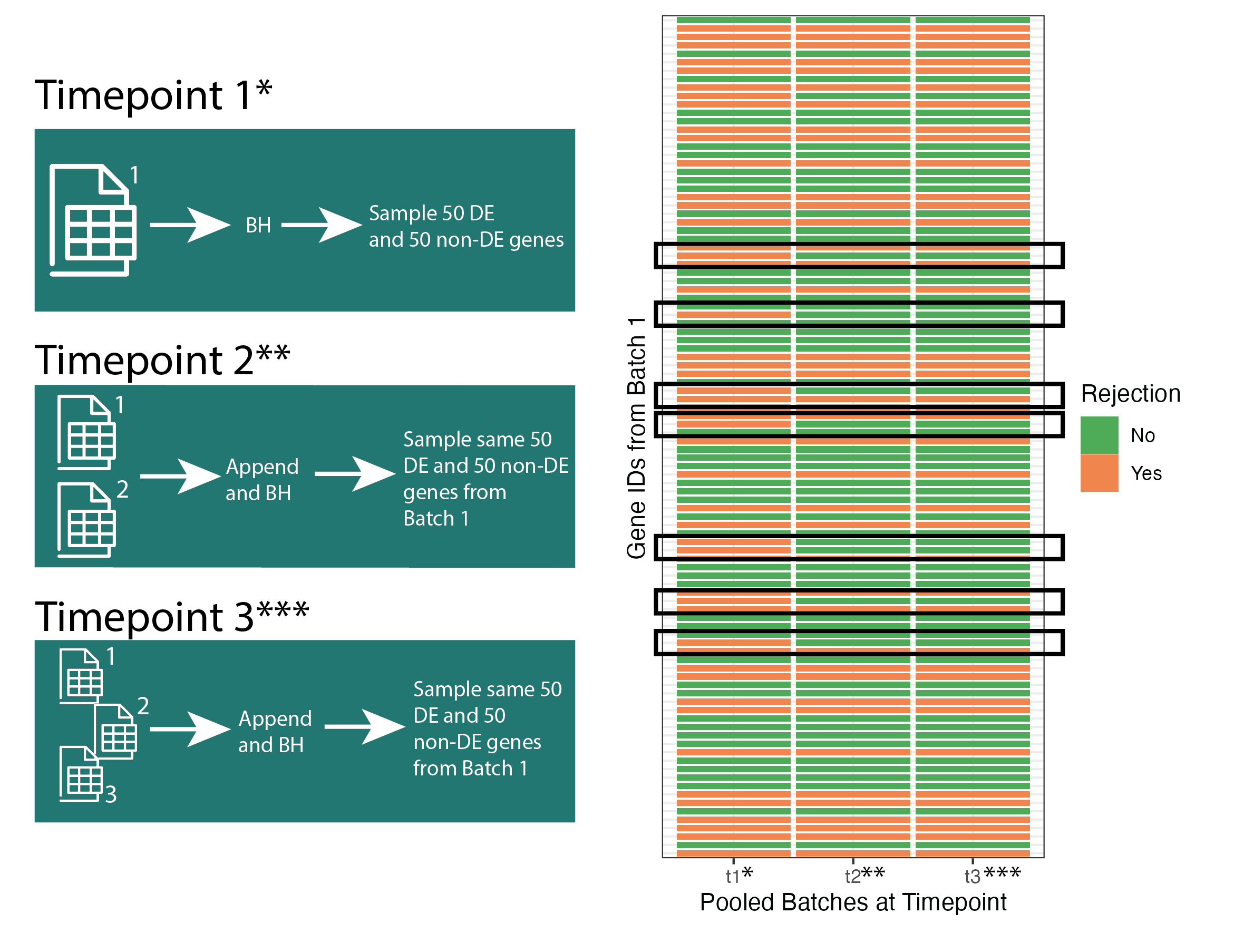}
\caption{Schematic visualizing the pooled offline approach and how gene differential expression results from Batch 1 are improperly changing (highlighted in black boxes). Online methods would produce a visualization akin to triplicating the t1 column.}
\end{figure}

\section{Simulation Study}

To explore the application of the online FDR control methods on a broader scale, we performed a RNAseq simulation study using the \textit{compcodeR} package \cite{soneson_compcoder--r_2014}, which simulates read counts from a negative binomial distribution (see Supplementary Appendix for more details). We used one of the standard practice workflows for RNAseq differential expression analysis and applied the \textit{voom} method \cite{law_voom_2014} from the \textit{limma} package (v3.50.0) \cite{ritchie_limma_2015}. 

In all our simulation experiments, 50 RNAseq datasets (referred to as ``batches") each with 10,000 genes were simulated to represent a group of batches arriving over time (which we henceforth refer to as a ``family"). For each family, we simulated a range of proportions of DE genes: 
\begin{gather*}
\{ 0.01, 0.02, 0.03, 0.04, 0.05, 0.06, 0.07, 0.08, 0.09, 0.1, 0.2, 0.3, 0.4, 0.5 \} 
\end{gather*}

We focus on two-group comparisons only. We specified 5 replicates for both the control and experimental groups, a log-fold change simulation parameter of 1.5, a sequencing depth of $10^7$, minimum and maximum factors that are multiplied with sequencing depth of $0.9$ and $1.1$ respectively, and a proportion of upregulated genes set to 0.5, which were all mostly default values in the \emph{compcodeR} package. We then ran \emph{limma-voom} as our differential expression analysis method. This simulation setup was replicated 100 times independently for computational reasons.

For each family, we applied offline BH, offline StBH, pooled offline BH, pooled offline StBH, and the online procedures BatchBH, BatchStBH, and BatchPRDS. The FDP was calculated as the number of false positives divided by the total number of genes declared DE. The average FDP across 100 trial runs provides empirical approximation of the true FDR. Power was calculated as the number of truly DE genes divided by the total number of genes declared DE and averaged over 100 trial runs. For offline BH and StBH, this entailed applying BH and StBH to each of the 50 datasets within a family, and then calculating FDR and power. For pooled BH and StBH this entailed applying BH and StBH a single time to all 50 datasets combined within a family. BatchBH, BatchStBH, and BatchPRDS were applied as described in \url{https://dsrobertson.github.io/onlineFDR/articles/onlineFDR.html} \cite{zrnic_power_2021}.


\begin{figure}[H]
\centering
\includegraphics[width=0.9\columnwidth]{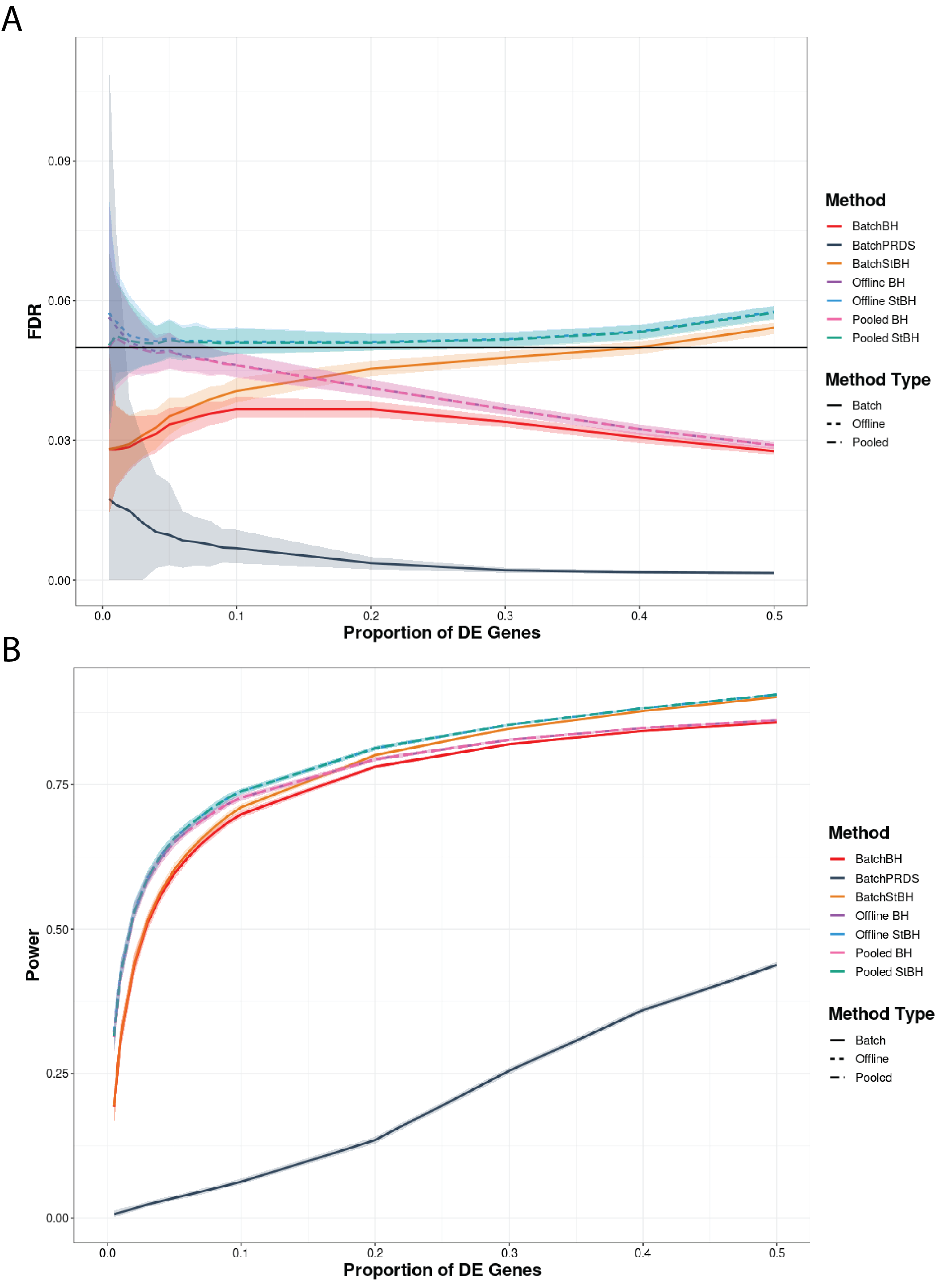}
\caption{FDR (A) and average power (B) versus proportion of DE genes ($\pi$) comparing offline, pooled offline, and online Batch algorithms at batch size of 10000. Number of batches set to 50 at a constant $\pi$. $\alpha$ set to 0.05. Simulated log fold-change set to 1.5. Shaded ribbons represent empirical 95\% confidence bounds.}
\end{figure}

Comparing the FDR across a range of proportions of DE genes the offline and pooled offline methods performed very similarly (Figure 4A). We observe that when either the offline BH (as is commonly done) or pooled offline BH approaches are performed on individual datasets, and the individual datasets are naively composed as a family of datasets arriving over time, the standard deviation of FDP is highly variable at low $\pi$ (e.g. $\pi \leq 0.1$), exceeding the pre-specified significance threshold. At higher $\pi$, FDR control is achieved by both offline BH and pooled offline BH. Offline StBH and pooled offline StBH on average slightly inflate the FDR above the threshold with a similarly high extent of variability at low $\pi$ as their BH counterparts. BatchBH and BatchPRDS control FDR well below the significance threshold throughout, with BatchPRDS exhibiting more conservatism. BatchPRDS is also highly variable at very low $\pi$; however, it still controls FDR at the nominal level as expected from its theoretical guarantees. In our simulation, BatchStBH experiences a slight inflation of FDR at higher $\pi \geq 0.4$. 

BatchBH and BatchStBH had close power to that of offline BH and pooled offline approaches starting from $\pi = 0.05$, which became closer as $\pi$ increased. BatchPRDS was very conservative across the range of $\pi$ tested. In this simulated scenario, all methods had low power at $\pi < 0.1$. Figure 4 shows that offline and pooled offline methods do not control the FDR in the online framework where online methods do, while still maintaining practically comparable levels of power.


\section{The effect of ordering}

Ordering of hypotheses can be crucial for the outcome of online testing, since it can lead to either a loss or gain of power \cite{javanmard_online_2018}. To increase power, hypotheses can be ordered, using side information, such that those most likely to be rejected are tested first. Hence, we sought to compare the performance of offline and online methods in a simulated ordered setting. 

We have simulated a `best-case' scenario in which a scientist decides to order RNAseq experiments so that batches with a higher proportion of DE genes are tested first, where we assume that there is \emph{a priori} justification. We represent this scenario by simulating the first 10 datasets (termed ``Super Batches") as having a higher proportion $\pi$ of DE genes, followed by 40 datasets (termed ``Regular Batches") that have a lower proportion $\pi$ of DE genes, such that the overall proportion $\pi$ of DE genes is one of $ \{ 0.1, 0.2, 0.3, 0.4, 0.5 \}$:

\begin{table}[h!]
\centering
\caption{Ordered Setup of Number and Proportion of Super Batches and Regular Batches}
\begin{tabular}{cccccc} \toprule
\textbf{Family} & \textbf{Super Batch} & \textbf{Super Batch $\pi$} & \textbf{Regular Batch} & \textbf{Regular Batch $\pi$} & \textbf{Overall $\pi$} \\
50              & 10                   & 0.3                     & 40                     & 0.050                      & 0.1                 \\
50              & 10                   & 0.3                     & 40                     & 0.175                     & 0.2                 \\
50              & 10                   & 0.5                     & 40                     & 0.250                      & 0.3                 \\
50              & 10                   & 0.5                     & 40                     & 0.375                     & 0.4                 \\
50              & 10                   & 0.5                     & 40                     & 0.500                       & 0.5                \\
\bottomrule
\end{tabular}
\end{table}

In this setting, offline StBH and pooled offline StBH do not control FDR at the specified level, across different $\pi$ (Figure 5A). The ordering of hypothesis tests has benefited the BH-based algorithms where both offline BH and pooled offline BH now control the FDR. BatchBH and BatchPRDS still control the FDR at various $\pi$; however, BatchStBH inflates FDR to comparable levels as $\pi > 0.2$. We return to this issue in the Discussion. The ordering has also benefited all approaches in terms of power. The power achieved by the online algorithms is very close to that achieved by their offline and pooled offline counterparts across all $\pi$.

\begin{figure}[H]
\centering
\includegraphics[width=0.9\columnwidth]{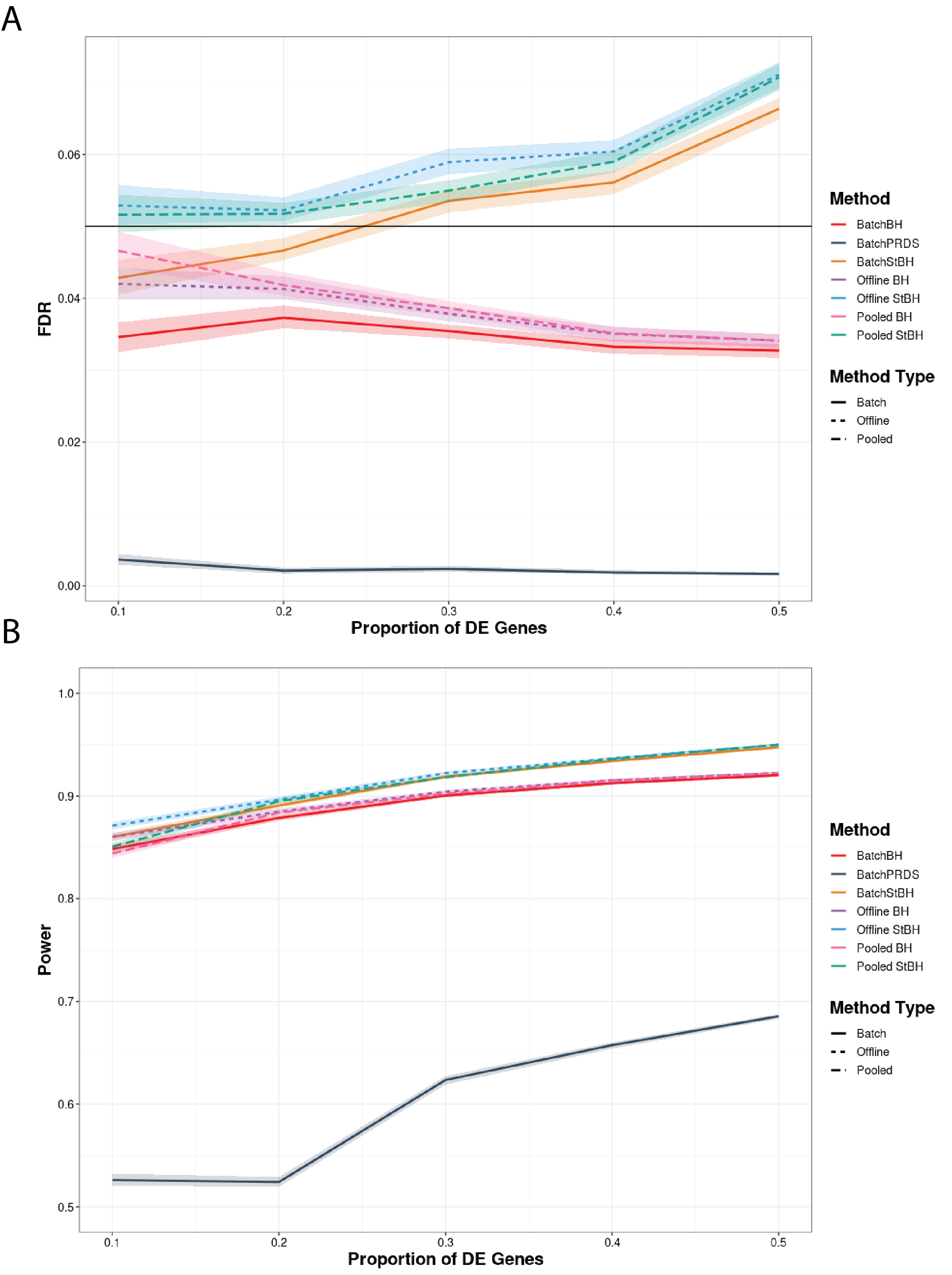}
\caption{FDR (A) and average power (B) versus probability of non-null hypotheses $\pi$ comparing offline, pooled offline, and online Batch algorithms at batch size of 10000 in an ordered setting such that datasets with a higher proportion of differentially expressed genes are tested first. Number of batches set to 50. $\alpha$ set to 0.05. Simulated log fold-change set to 1.5. Shaded ribbons represent empirical 95\% confidence bounds}
\end{figure}

\section{Discussion}

In this paper, we demonstrate that the online FDR control algorithms provide a principled way to guarantee control of FDR at a nominally specified level in an online paradigm, e.g. RNAseq datasets arriving sequentially over time, and at every time step, a batch of decisions is made via the BH or StBH algorithm. We show that offline approaches have variable control of FDR of related RNAseq experiments over time in certain settings simulations, and that a pooled application may improperly change previous differential expression rejection decisions, which can be construed as one form of p-hacking \cite{head_extent_2015}. Furthermore, in some simulation scenarios, we observe empirically that online approaches have comparable power to offline approaches.

Using three real RNAseq datasets that we considered as a ``family" of datasets arriving over time, we applied the online FDR control methodologies. Although our datasets were limited in that we could not assess ``ground truth", we see that the online methodologies are broadly more conservative than their offline and pooled offline counterparts, which is likely driven by the theoretical long-term FDR control conferred. Notably, as the number of datasets increase, the online methods are able to recapitulate a higher proportion of genes as detected by more conventional offline methods, which suggests that while the online approaches may have lower statistical power in the short-term, in the long-term, they may have practically comparable power to offline methods. We selected the datasets that were similar enough in terms of identical time points, treatment, and mouse model to reduce measurable confounding to the best of our ability. But we also remark that the primary motivation behind this analysis was not biologically driven, in that questions such as whether similar gene biology was detected across the datasets were considered to be beyond the scope of this paper. Generally, similar RNAseq datasets added to a growing data repository such as the Gene Expression Omnibus (GEO) are potentially indicated use cases for the application of online FDR algorithms. 

In our unordered simulation setting, we observed that the online algorithms successfully controlled FDR across a wide range of $\pi$. Is is notable that particularly at low proportions of differentially expressed genes, the offline approaches did not maintain control of FDR. The range of $\pi$ that we test may be relevant for the range of number of DE genes that may be detected in practice \cite{corley_differentially_2017}. BatchStBH had slight inflation of FDR, but that was at a very high $\pi$ of 0.5 in which the distribution of DE and non-DE genes in a RNAseq dataset is symmetrical, which is likely very rare in practice. The slight inflation of FDR could have also resulted from the subtle dependency introduced in our simulation methods. The \emph{compcodeR} package draws $\mu$ and $\phi$ estimates for read count generation from the same underlying real-world data distribution, so it is possible that, although we have simulated the datasets independently to the best of our ability, there still exists some degree of dependency amongst the \emph{p}-values. BatchPRDS was more conservative than BatchBH and BatchStBH; it guarantees FDR control when \emph{p-}values are dependent within each batch but at the cost of a substantially lower power than any of the other algorithms. In real RNAseq datasets, it is likely that DE gene \emph{p-}values are dependent \cite{subramanian_gene_2005}. Future work should investigate the performance of Batch online FDR control methods within varying dependency structures. Subramanian et al. also proposed evaluating sequencing data at the level of gene sets since single-gene analysis may miss important effects on pathways. Thus, one other potential future research direction is to apply online control methods to gene set analysis \emph{p-}values.

We also simulated an ordered setting, since in practice, scientists may order batches of hypotheses in such a way that seeks to optimizes their power and reflects their \emph{a priori} knowledge. We show in our simulations that ordering hypotheses increases the power of online methods, but appears to inflate the FDR of BatchStBH at higher $\pi$. We note a limitation of our simulation where in practical scenarios, it may be a more nuanced decision between ordering batches of RNAseq datasets or the hypotheses (i.e. gene \emph{p}-values) themselves. In our application of online methods to our real-world data, we did not explore \emph{a priori} ordering of datasets as biological rationale was not the emphasis. However, it is possible that if we were to test Batch 2 after both Batches 1 and 3 (as it appeared inherently less powerful) for some valid biological justification (e.g.\ this was a technical pilot of the RNAseq experimental setup itself where high signal was not expected), we may achieve better overall power across the three datasets together in an online framework. Future studies could investigate ordering retrospectively (e.g.\ in a database such as GEO) or prospectively (e.g.\ making decisions how to order datasets arriving over time). We reiterate that the purpose of our simulation was to represent how much more powerful online methodologies could be if we happened to have inherently more powerful datasets ordered first. In addition, in reality, we may not expect such a high proportion of DE genes, nevertheless we caution the use of BatchStBH in an ordered setting as it might ``greedily" increase statistical power at the expense of FDR.

%
Additional future directions of exploring FDR control with RNAseq data can explore and benchmark how varying batch size, the number of samples, and using different gene filtering methods or differential expression methods would affect performance of online algorithms. Using other FDR control methods (besides BH and Storey-BH) interpolated in the online manner as we have described is also possible. 
\\
    
\section*{Acknowledgements}

We would like to thank Eric Muise, Wendy Blumenschein, Doug Linn for their permission to use the real-world datasets as well as their comments and feedback. We would also like to thank Ola Olow and Uwe Mueller for their detailed comments and suggestions to improve this manuscript.

\section*{Conflicts of Interest}

We have no conflicts of interest to declare.

\section*{Funding}

This work was supported by the UK Medical Research Council [MC\_UU\_00002/14 to D.S.R.], the Biometrika Trust [to D.S.R.] and the NIHR Cambridge Biomedical Research Centre [BRC1215-20014 to D.S.R.]. The views expressed in this publication are those of the authors and not necessarily those of the NHS, the National Institute for Health Research or the Department of Health and Social Care (DHSC). For the purpose of open access, the author has applied a Creative Commons Attribution (CC BY) licence to any Author Accepted Manuscript version arising.

\section*{Data availability statement}

Code used for the simulation studies can be found at \url{https://github.com/latlio/onlinefdr_rnaseq_simulation}. The real-world RNAseq data underlying this article will be shared on reasonable request by the corresponding author.

\printbibliography

\end{document}


\maketitle

\onehalfspace

\section{Supplementary Appendix}

\subsection{Intuitive sketch of why pooling datasets does not control FDR}

For simplicity, consider a situation of testing $2n$ hypotheses from only two batches (the argument holds for multiple batches as well). Suppose we are testing $2n$ hypotheses with BH (or StBH). The false discovery rate controlled using the BH procedure is defined as:
$$FDR = E[Q] = E \left[ \frac{V}{R} \right] \leq \alpha$$
where $Q$ is the false discovery proportion, $V$ is a random variable representing the number of false rejections from the $2n$ hypotheses, $R$ is a random variable representing the total number of rejections from the the $2n$ hypotheses, and $\alpha$ is the nominal FDR control rate.

Now suppose that we split these $2n$ hypotheses into two batches of size $n$ each. Let $R_1$ and $R_2$ denote the number of rejections in batch 1 and 2, respectively, and $V_1$ and $V_2$ denote the corresponding number of false rejections. We then define $Q_1 = \frac{V_1}{R_1}$ and $Q_2 = \frac{V_2}{R_2}$. The overall false discovery proportion using this naively composed BH procedure is given by:
$$Q' = \frac{V_1 + V_2}{R_1 + R_2}$$
It follows that:
$$E[Q'] = \frac{V_1}{R_1+R_2} + \frac{V_2}{R_1+R_2} \leq \frac{V_1}{R_1} + \frac{V_2}{R_2}$$
We know that $E[Q_1] = E[\frac{V_1}{R_1}] \leq \alpha$ and $E[Q_2] = E[\frac{V_2}{R_2}] \leq \alpha$, and hence:
$$E[Q'] \leq E \left[ \frac{V_1}{R_1} \right] + E \left[ \frac{V_2}{R_2} \right] \leq 2\alpha$$
This gives us an upper bound on the FDR, but no guarantees on the lower bound.

Now suppose that the proportion of non-nulls is 0, and imagine repeatedly generating \emph{p}-values. We could quite conceivably see something like this for 20 simulation replicates: for 18 replicates $R_1 = R_2 = 0$ (and hence $V_1 = V_2 = 0$), for one replicate $R_1 = V_1 = 1$ and $R_2 = V_2 = 0$, and for one replicate $R_1 = V_1 = 0$ and $R_2 = V_2 = 1$. Hence $E[Q_1] = E[Q_2] = \frac{1}{20}$ as expected, but $E[Q'] = \frac{2}{20}$. If we were to increase 20 replicates to $10^5$ replicates for example, this effect would be substantially diluted (i.e.\ $E[Q']$ would be far lower than the upper bound of $2\alpha$), but this gives intuition as to why FDR can be inflated at low proportions of non-nulls. 

\subsection{Details of Methods}

In both the real-world application and the simulation study, we set the $\gamma_i$ sequences to $\{0.5, 0.5, \ldots, 0\}$ for BatchBH and BatchPRDS as recommended by Zrnic et al. (\cite{zrnic_power_2021}) for larger batch sizes. We used the default $\lambda = 0.5$ for BatchStBH. We use the default $\gamma_i$ sequence for BatchPRDS.

For our simulation setup, we relied on previous literature to decide RNAseq synthetic data settings. We selected 10,000 genes, 5 samples, and our fold change value of 1.5 following previous RNAseq simulation literature (\cite{li_sample_2019}, \cite{lai_statistical_2017}, \cite{yu_power_2017}, \cite{bi_sample_2016}, \cite{soneson_comparison_2013}). Low sample size has been demonstrated to adversely affect FDR control of differential expression analysis in general (\cite{soneson_comparison_2013}). We selected our sequence of $\pi$ following the online FDR literature (\cite{zrnic_power_2021}, \cite{robertson_onlinefdr_2019}), although in practice, we don't expect the the proportion of truly DE genes to be as high as 0.5. We also set the proportion of DE genes regulated in either the up direction or down direction to 0.5, as Soneson et al. showed that when DE genes were not regulated in different directions, the ability to control the FDR was heavily impaired (\cite{soneson_comparison_2013}). We simulated the data using \emph{compcodeR}, which is an R package designed as a tool to benchmark new approaches on synthetic RNAseq data (\cite{soneson_compcoder--r_2014}). We used \emph{limma-voom} as our differential expression method since it is a widely-used method that generates reasonable results (\cite{soneson_comparison_2013}).

In the limma-voom workflow, we excluded genes using the \emph{edgeR} method, which is described by Chen et al. (2016)~\cite{chen_reads_2016}.  Removing low count genes may allow for more genes to be declared DE and reduce the computational expense of subsequent analysis. We also use the trimmed mean of $M$ values (TMM) method as described by Robinson and Oshlack~\cite{robinson_scaling_2010} as our count normalization method.

All code used for this analysis can be found at: \url{https://github.com/latlio/onlinefdr_rnaseq_simulation}

\printbibliography